# Effects of Thermal Modification on the Flexure Properties, Fracture Energy, and Hardness of Western Hemlock

T. Nakagawa[a], E. Poulin[b], T. Rueppel[c], Z. Chen[a], J. Swinea[c], M. O'Brien[b], G. Houser[b], G. Wood[b], M. Weinheimer[e] P. Bahmani[d], P. Stynoski[c], M. Salviato[a]*

[a]William E. Boeing Department of Aeronautics and Astronautics, University of Washington, Guggenheim Hall, Seattle, WA, 98195-2400, USA

[b]Composites Recycling Technology Center (CRTC), 2220 W 18th St, Port Angeles, WA 98363-1521, USA

[c]Engineering Research and Development Center (ERDC) Construction and Engineering Research Laboratory (CERL),2902 Newmark Drive, Champaign, IL 61822, USA

[d]Department of Civil and Environmental Engineering, Washington State University, 405 Spokane St, Sloan 101, Pullman, WA 99164-2910, USA

[e]Chickadee Forestry LLC, 2023 E Sims Way #147, Port Townsend, WA 98368-6905 USA

**ABSTRACT**

This study investigates the effect of thermal modification on the flexural properties, transverse fracture energy, and hardness of western hemlock, a material which is finding increasing applications in construction. Flexure tests on specimens featuring longitudinal and transverse grains showed that thermal modification at 167˚C slightly improves the flexural modulus and strength and leads to less statistical variability compared to unmodified samples. On the other hand, the fracture and Janka hardness tests revealed a more pronounced brittleness of the thermally modified samples. In fact, the total mode I fracture energy of modified Single Edge Notch Bending (SENB) samples was about 47% lower for radial-longitudinal systems and 60% lower for tangential-longitudinal systems. Similarly, the average Janka hardness in the tangential, radial, and transverse planes was 8.5%, 3.9%, and 9.4% lower in the modified specimens, respectively.

The results presented in this work show that thermal modification can have a significant effect on the fracturing behavior of western hemlock and its energy dissipation capabilities. For design, this must be taken into serious consideration as these properties significantly influence the damage tolerance of this wood in the presence of stress concentrations such as e.g., those induced in bolted joints and cut outs. Fracture energy and hardness are also strongly correlated to ballistic performance.

## 1 Introduction

Wood is a widely used natural material in various construction and design applications due to its desirable mechanical properties and aesthetic appearance (Fridley 2002). However, wood is also prone to deterioration over time due to environmental factors such as moisture,

*Address all correspondence to this author. Email: salviato@aa.washington.edu

UV radiation, and biological attack (Reinprecht 2016). To enhance the durability of wood and increase its resistance to decay, thermal modification has emerged as a promising treatment method. This process involves subjecting wood to high temperatures, usually between 160 and 240°C, in the absence of oxygen, which causes changes in its chemical and physical structure (Hill, Altgen and Rautkari 2021, Militz and Altgen 2014, Sandberg and Kutnar 2016). These changes lead to improvements in properties such as dimensional stability, and microbial resistance while also reducing wood's moisture absorption (Hill, Altgen and Rautkari 2021). Several studies have investigated the effects of thermal modification on wood mechanical properties and durability showing that results can vary depending on the species of wood, temperature and duration of the treatment, as well as other factors (Bourgois and Guyonnet 1988, Hillis 1984, Kubojima, Okano and Ohta 2000, Lekounougou, et al. 2011).

Hakkou et al. reported that thermal modification of beech wood increased its dimensional stability, hardness, and decay resistance (Hakkou, et al. 2006). Similarly, a study by Yildiz et al. showed that thermal modification of Spruce wood resulted in improved dimensional stability and resistance to fungal decay (Yildiz, Gezer and Yildiz 2006).

Pleschberger et al. (Pleschberger, et al. 2014) studied the fracture behavior of ash modified at a temperature of 200, 210, and 220 °C in radial/longitudinal and tangential/longitudinal direction at 65% air relative humidity (RH). They reported an increase in brittleness of the material along with a reduction in fracture energy. Similar results were also reported on the same material by Majano-Majano et al. (Majano-Majano, Hughes and Fernandez-Cabo 2010) and on spruce wood by Murata et al. (Murata, Watanabe and Nakano 2013).

Standfest and Zimmer (G. and B. 2008), reported an increase in Brinell hardness of ash in the longitudinal direction while in the tangential and radial direction the hardness decreased. On the other hand, Govorčin et al. (S., T. and R. 2009) showed a reduction of ash hardness in the principal anatomical directions after heat treatment at a temperature of 200 °C. They also reported a decrease in Modulus of Rupture (MOR) and compression strength in the longitudinal direction.

Roszyk et al (Roszyk, et al. 2020) investigated the moisture-dependent strength anisotropy of thermally modified European ash in compression. They showed that thermal treatment kept the intrinsic anisotropy of wood mechanical properties. It decreased wood hygroscopicity, which resulted in improved strength and elasticity measured for wet wood when compared to untreated and treated samples.

Nhacila et al. (Nhacila, et al. 2020) studied the effects of thermal modification on the physical and mechanical properties of Mozambican *Brachystegia spiciformis* and *Julbernardia globiflora* wood. For B. *spiciformis*, they showed that the Modulus of Elasticity (MOE) decreased by 10.2%, the Modulus of Rupture (MOR) by 50.8%, compression strength parallel to the grain by 29.2% and Brinell hardness by 23.5%. Timber of *J. globiflora* followed the same trend with an MOE decrease by 6.9%, an MOR decrease by 53.2% and a decrease in compression strength parallel to the grain by 21.9%.

Boonstra et al. (Boonstra, Van Ackerb and Tjeerdsmac 2007) investigated the effects of heat treatment on a number of softwoods including Radiata pine, Scots pine, and Norway spruce. They showed that, in general, heat treatment in these softwoods leads to a large decrease in the tensile strength parallel to the grain whereas the compressive strength parallel to the fibers increased. They also found a quite significant reduction in impact strength.

While several studies have been focused on the effect of heat treatment on a number of wood species, far less attention has been devoted to the study of western hemlock, especially when it comes to its fracture energy and hardness which are important indicators of its damage tolerance and ballistic performance. A preliminary study has been published recently by Nourian and Avramidis (Nourian and Avramidis 2021) who investigated the effects of commercial thermal modification on western hemlock by performing evaluations of basic density, hygroscopicity, water absorption, anti-swelling efficiency, color change, Janka hardness, and dynamic modulus of elasticity. Their results revealed that basic density, hygroscopicity, and water absorption decreased at higher treatment temperatures, while dimensional stability considerably increased. On the other hand, the mechanical behavior was not significantly affected by the thermal treatment. However, the results did not cover the fracture behavior which is an important aspect for design with this type of wood. The goal of the present article is to take a step in filling this knowledge gap by providing an extensive investigation on the effect of thermal modification on the longitudinal and transverse flexural behavior, fracture energy, and hardness in western hemlock.

Thermal modification offers some potential tremendous benefits to building envelopes and building science. But with the low volume of available thermally modified lumber worldwide, producers have typically focused on value applications and not made serious attempts towards structural applications. In the course of this development at the Composite Recycling Technology Center (CRTC) but outside the detail of this study, CLT panels were manufactured from thermally modified coastal western Hemlock (Tsuga Heterophylla) and machined to final dimension of .86m width and 3.35m in length, with a square-edge tongue and groove (t&g) feature along the long-edges. The clearances for the side of the t&g interlock were net 1mm, so equivalent to .5mm between each face of the tongue to groove with no taper applied. The CNC machined panels were stored indoors but in an uncontrolled environment with temperatures varying between 7ºC and 30ºC, and relative humidity estimated to vary between 50% and 80%. Storage time was 6 to 9 months for these panels. Coastal western hemlock typically has significant issues in dimensional stability caused by internal stresses, inconsistent drying and pockets of moisture (Song 2019), and exhibits warping, cupping, and twisting on milling to final lumber. Application to cross-laminated timber (CLT) is limited due to these process-induced defects, and tight framing is quite difficult to achieve. One would expect, after precise machining and an extended storage time with varying environmental conditions 0.5mm of clearance would be problematic, but the results were quite encouraging.

After this storage duration a 22 sqm demonstration structure was assembled at the CRTC with 22 interlocking t&g panels. The dimensional stability of the thermally modified coastal western hemlock (CWH) was such that panels slid together without interference or any coercion required. This structure is shown in Figure 1. It's unlikely that this would have been possible with traditional CWH lumber, even kiln-dried, as hygrothermal dimensional changes in the overall panel as well as in the t&g clearance dimensions would have caused misalignment resulting in either binding or loose fit. This capability enables a tight and durable building envelope seal, as well as a consistent glue line thickness and higher-performance assembly.

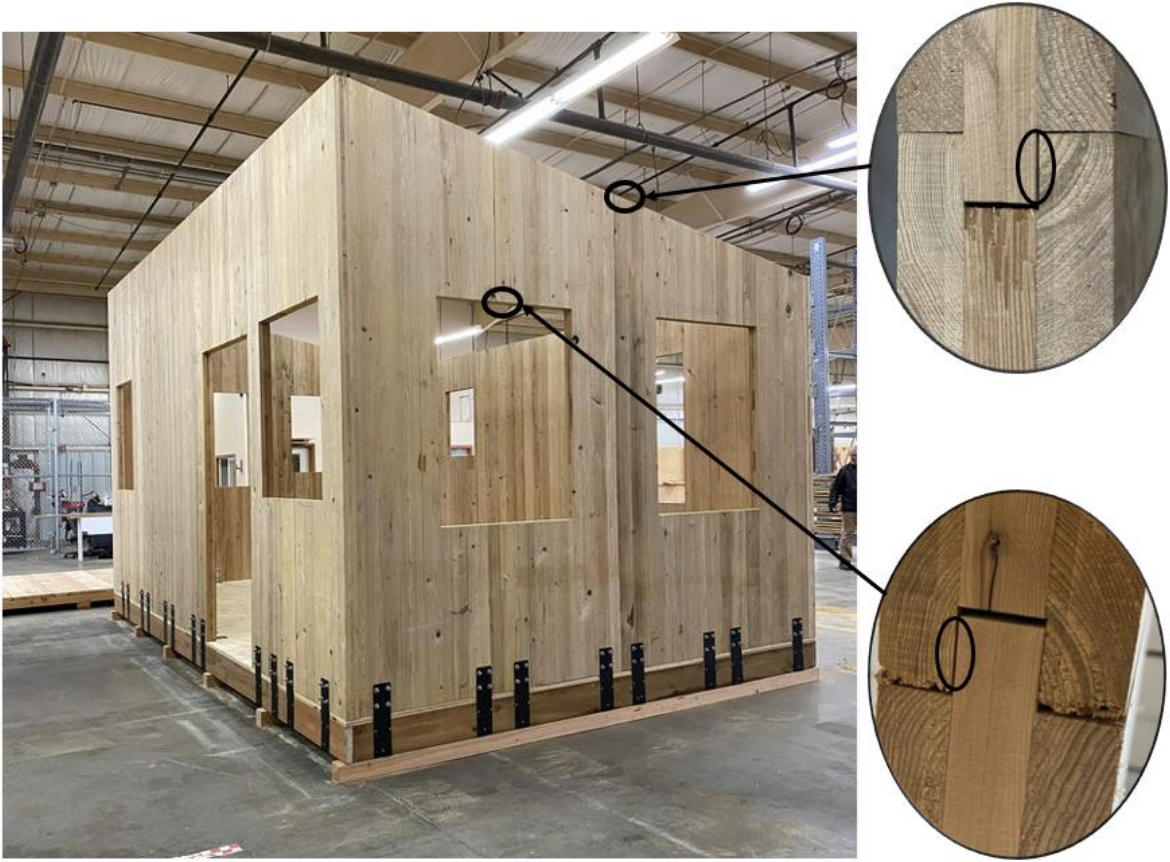

Figure 1: 22 sqm demonstration structure machined and assembled at the CRTC with thermally modified CWH CLT panels. Panels were manufactured at the Composite Materials & Engineering Center at WSU. Note that the tight 1mm of clearance is between the side faces of the t&g joint (circled) to promote panel alignment. A larger gap was intentionally designed between the end faces of the t&g joint to account for assembly tolerances.

The results of the present article not only provides useful design guidelines to account for the effect of thermal modification on the mechanical behavior of western coastal hemlock, but also represents a first step towards the construction of rich databases to calibrate and validate computational design tools.

This research seeks to act as a reference for integrating a thermally modified undervalued US based wood species into structural design standards and computational models. Advances in the structural relevance of thermally modified wood through the complete understanding of mechanical behavior would positively impact the use of the material in mass timber elements like CLT.

## 2 Materials and Methods

### 2.1 Materials and Preparation

Sample material for this study was acquired from a small harvest of coastal Western Hemlock from the Makah reservation, which is located on the northwestern tip of the Olympic Peninsula in Washington state. The forest is designated as a site class 3, low elevation, with a mean annual air temperature of 8.89°C. The forest is characteristic for a pacific northwest climate with large amounts or rainfall in the winter months, and a mean annual precipitation of 203.2 -304.8 cm. This timber was grown intentionally as part of a commercial timber harvest program, and the forest was managed for productive growth. A 55.88 cm diameter, 396 cm log was milled with a Weyerhaeuser mobile sawmill (Weyerhaeuser inc. 2023) into a variety of board sizes, to meet various testing needs.

For this study, two 3.175×40.64×396 cm boards were cut tangentially from the middle section of the log's radius. All longitudinal and transverse bending specimens were acquired from the outermost board, and all longitudinal and transverse Single Edge Notch Bend (SENB) specimens were acquired from the board adjacent. One 6.35×18×396 cm board was cut tangentially near the pith of the tree for samples to test Janka hardness. Figure 2 shows the location of the two boards designated for these testing specimens.

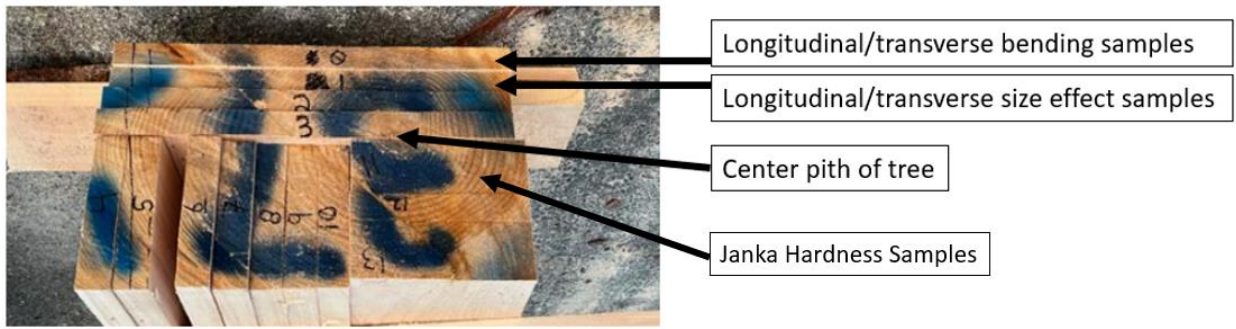

Figure 2: Test sample board locations.

Each 396 cm board was cut into 3 equal 132 cm lengths. All 6 lengths were conditioned at approximately 21.11°C and 65% relative humidity for 2 weeks before an accelerated drying regime. Due to time constraints, the wood was dried in a Wisconsin oven for up to 4 days with temperatures not exceeding 48.89°C. Moisture content measurements were taken daily on a freshly planed surface with an Orion 930 moisture meter. Boards were removed from the drying regime once their moisture content dropped below 13%. Some cracking and shrinkage effects were observed, but effects were minimal, with the ability to produce straight crack free samples unaffected. One length of each bending, SENB, and Janka hardness board was set aside and allowed to remain at ambient conditions until the time of individual specimen preparation. These boards constitute the un-modified (UM) samples.

After drying, the other two lengths of bending and size effect boards were sent to Therma Wood Technologies in Poulson, Montana for a thermal modification treatment. The wood underwent a standard production cycle, with a 7 ½ hour total run time. Temperature was increased to 167.78°C over 3 ½ hours, dwelled at 167.78°C for 1 ½ hours, and then cooled to 70°C. The pressure profile is proprietary but designed to complement the thermal cycle and expedite the modification process. These boards constitute the thermally modified (TM) samples.

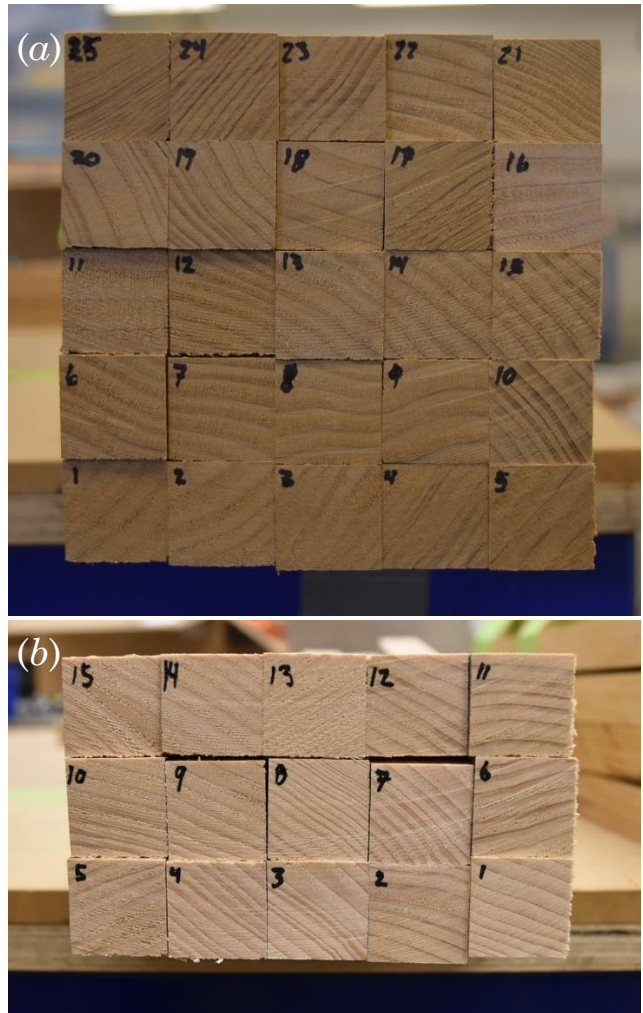

Figure 3: Grain structure of (*a*) TM and (*b*) UM longitudinal bending specimens. Loading direction is down from the top of the image.

## *2.2 Specimen Preparation*

### *2.2.1 Longitudinal and Transverse Bending Specimens*

Longitudinal and transverse bending specimens of UM and TM wood were prepared as per ASTM D143 (ASTM D143 2022) for secondary method specimens. Boards were planned on a Grizzly planar down to 25 mm thickness. Then, 25×25×410mm longitudinal specimens were ripped parallel to the grain on a Sawstop table saw, sampled from the edge farthest from the pith of the tree. They were run back through the planar at 25 mm to reduce dimensional variability from the table saw prior to being cut to length. Transverse specimens spanned the entire width of the board and were ripped perpendicular to the grain. All samples were cut to length on a Dewalt miter saw. Cross sections of grain structure for UM and TM longitudinal bending samples are shown in Figure 3. 25 UM and TM samples were produced for each grain orientation.

### *2.2.2 Single Edge Notch Bending (SENB) specimens*

SENB specimens were prepared for fracture testing in a similar fashion to bending specimens. The planar was used to achieve more precise dimensions whenever possible. The edge notch was prepared by marking out the length of the crack and using a small kerf, 0.254 mm thick blade hand saw run through a miter box to cut the crack. It is worth noting that while this implies that the initial crack has a finite width, extensive research has shown that this does not affect the fracture behavior provided that the crack tip radius is smaller than Irwin's characteristic length (Salviato, et al. 2016, Bazant, Le and Salviato 2021, Ko, Davey, et al. 2019, Ko, Yang, et al. 2019, Li, et al. 2021, Qiao and Salviato 2019, Kumagai, et al. 2020). This will be the case in the following sections.

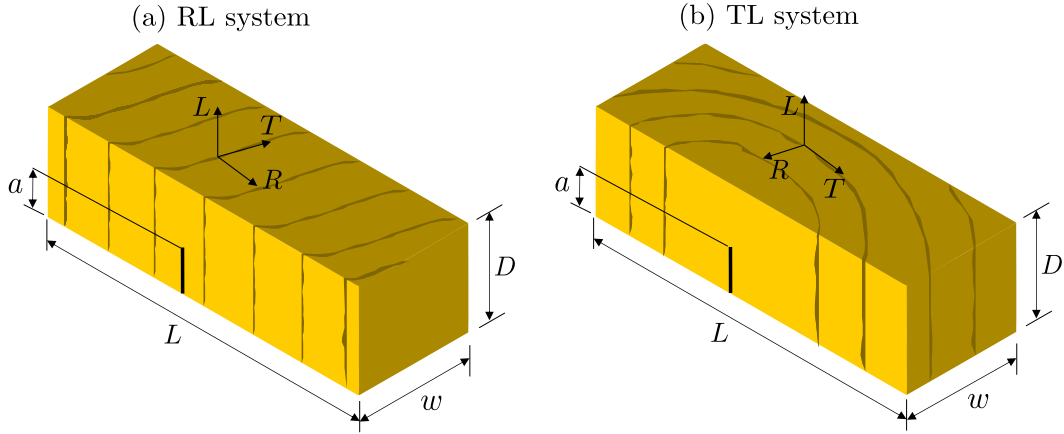

Figure 4: Configurations considered for the Single Edge Notch Bending (SENB) tests: (a) Radial-Longitudinal (RL) system and (b) Tangential-Longitudinal (TL) system. Dimensions are reported in Table 1.

|  |  | $D$ [mm] | $w$ [mm] | $L$ [mm] | $a$ [mm] |
|---|---|---|---|---|---|
| **RL system** | Size 1 | 12 | 20 | 80 | 6 |
|  | Size 2 | 24 | 20 | 160 | 12 |
| **TL system** |  | 36 | 20 | 240 | 18 |

Table 1. Dimensions of the SENB specimens for the two configurations investigated.

For the fracture tests, two configurations were tested. As shown in Figure 4, the first configuration, called Radial-Longitudinal (RL), sees the crack propagating parallel to the grain in plane LT. The second configuration, called Tangential-Longitudinal (TL), sees the crack propagating parallel to the grain in plane RL. Since the micro/mesostructure in front of the crack tip is different for the two configurations, the fracture energies are going to be different. Hence both configurations were tested. 8 samples were produced for each size and configuration.

Previous investigations on spruce woods (Murata, Watanabe and Nakano 2013) showed that the RL system typically features larger fracture energies and Fracture Process Zones (FPZs) compared to the TL system. When the size of the FPZ is not negligible compared to the structure size, size effect occur which might lead to the measurement of size dependent fracture energy if not properly accounted for (Bazant, Le and Salviato 2021). To make sure that the FPZ was fully developed and a size independent fracture energy could be measured, two sizes of RL system were tested. Since this system was the most prone to size effects, once it was verified that the measured fracture energy was not affected by size it was decided to test only one size for the TL system. A summary of all the specimen sizes is provided in Table 1 while Figure 4 shows a schematic representation of the configurations investigated in this work.

Dimension and weight measurements were taken, and samples were speckled for digital image correlation. Average moisture content for UM wood was 12.5% and average moisture content for TM wood was 9%.

### 2.2.3 Janka hardness specimens
Twenty UM and twenty TM 50.8×50.8×27 mm wood blocks were cut from the same log for Janka hardness testing, as shown in Figure 5. The annual growth ring patterns were very similar between the two configurations and each specimen exhibited 7-8 rings having wide earlywood and thin latewood bands. The densities of the specimen were obtained at Equilibrium Moisture Content (EMC) prior to testing.

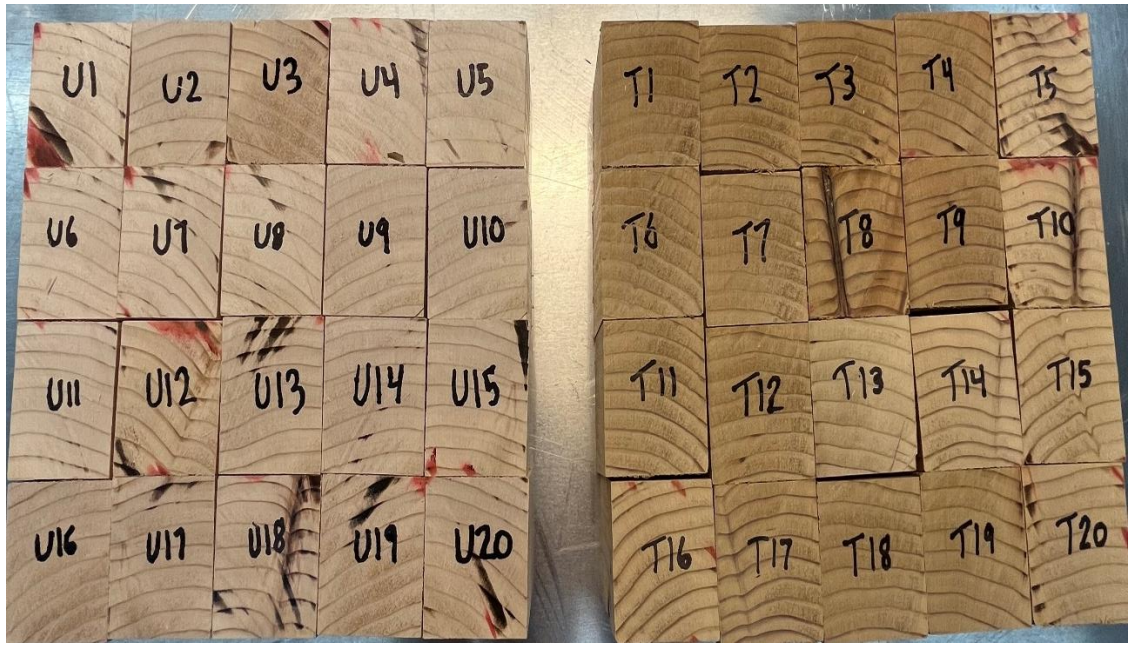

Figure 5: UM (left) and TM (right) 50.8×50.8×27mm Western hemlock blocks for Janka hardness testing.

## 2.3 Testing

### 2.3.1 Flexure and fracture tests
The flexure and the Single Edge Notch Bending (SENB) tests were performed on a Test Resources 316 series UTM with a 22kN load cell with a sampling rate of 10Hz (the typical test setup is shown in Figure 6). A displacement rate of 5.08 mm/min was utilized during the tests. Samples were prepared for Digital Image Correlation (DIC) with a thin coat of spray paint primer and fine random distribution of black speckles using black spray paint. Images were taken with a Nikon D5600 DSLR camera with a Nikon DX VR lens and a sampling rate of 1Hz. Thanks to DIC it was possible to characterize the whole strain field and strain redistributions at damage locations. Through DIC it was also possible to estimate the compliance of the fixture and the machine and verify that the crosshead displacement measured by the load frame was sufficient for deflection measurements.

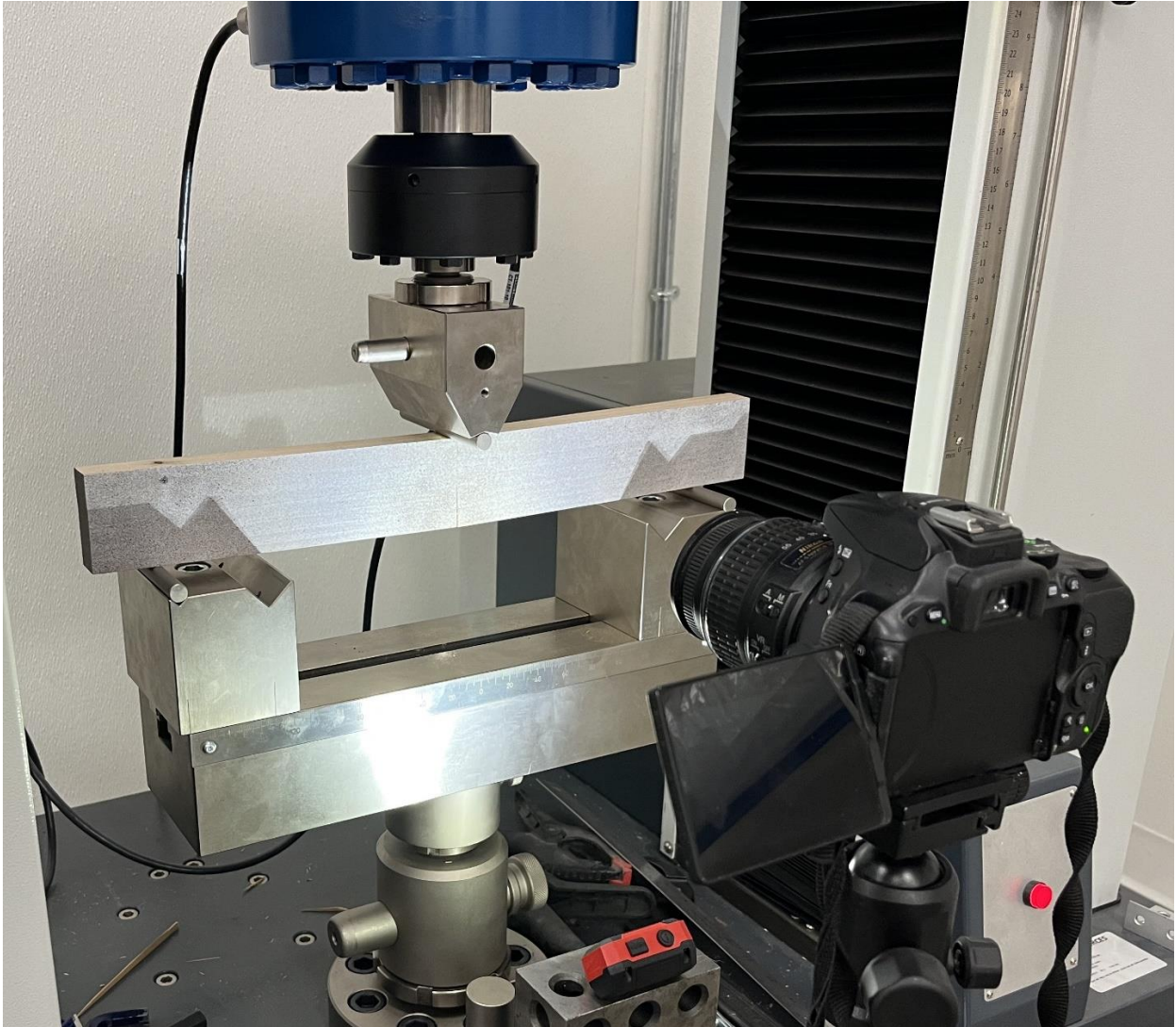

Figure 6: Test setup for SENB and longitudinal/transverse bending testing. SENB tests used 10mm rollers (shown) and longitudinal/transverse bending used 30mm rollers.

### *2.3.2 SEM Imaging*
After fracture tests were completed, one RL and one TL specimen of the two configurations were taken for SEM imaging of the fracture surface. A Phantom ProX SEM was used with an electron beam voltage of 15 kV and a chamber pressure of 60 Pa. A razor blade was used to cut sections of the fractured surface. Images taken were 5-15 mm away from the crack time and at least 1 mm way from the edges to make sure no damage seen in the images are from the sample prep.

### *2.3.3 Janka hardness tests*
A variation of a Janka hardness indenter was machined to a 11.3 mm diameter according to ASTM D143 (ASTM D143 2022). A drawing of the indenter is shown in Figure 7 and Figure 8 is the fixture resting on a wood block. The Janka hardness indenter was fastened to a United SFM-300KN Electro-Mechanical Series Universal Testing Machine and data was recorded via the United Datum5i software. The specimen was loaded at 6.35 mm/min to a maximum specimen penetration depth of 5.65 mm (half the diameter of the fixture). Six hardness values were obtained from each of the 40 specimens – two from the tangential face, two from the radial face, and two from the transverse face or cross section of the wood. Average values were then calculated for each face of each sample and ultimately each face for each of the two configurations.

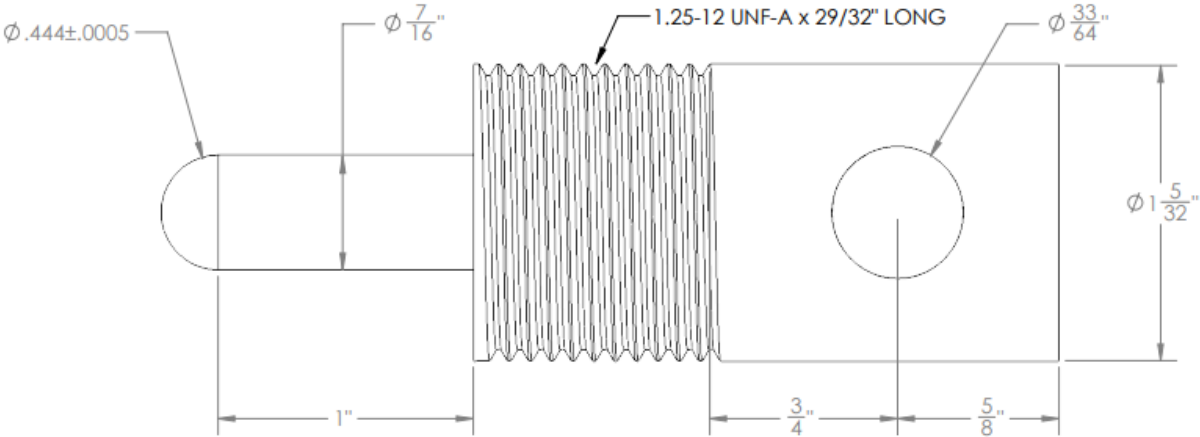

Figure 7: Machined Janka hardness indenter.

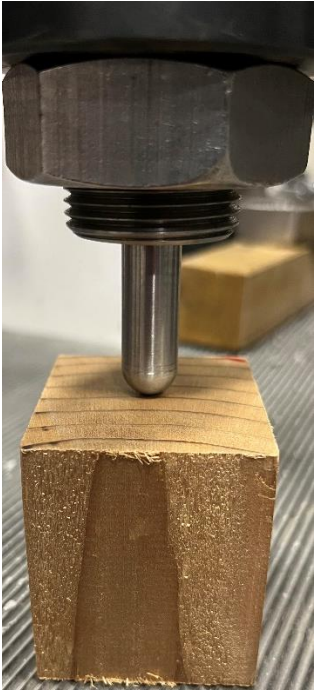

Figure 8: Janka hardness indenter on specimen.

# 3 Results and Discussion

## 3.1 Longitudinal and Transverse Bending Tests

### 3.1.1 Failure modes

Unmodified (UM) longitudinal bending samples failed in tension, although they exhibited several types of tensile failures. In accordance with the types of static bending failures described in ASTM D143 (ASTM D143 2022), UM longitudinal samples mostly failed in simple and splintering tension except for a few which failed in a cross-grain manner. Thermally Modified (TM) longitudinal bending samples failed in similar fashion according to the ASTM D143 failure types. It's notable that although the results of the bending test produced visually similar failures, the UM wood failures were typically slower and characterized by larger deformations compared to the TM wood. Part of the increased deformation in the UM samples was from crushing at the loading head. The TM wood failures were quicker, failing completely rather than slowly cracking and re-loading as seen with the UM wood. Additionally, there was less crushing in the TM wood at the loading head. In general, the TM wood behaved in a more brittle manner, while the UM wood withstood more deformation and cracking prior to ultimate failure. Figure 9 shows a typical fracture surface of the longitudinal flexure specimen.

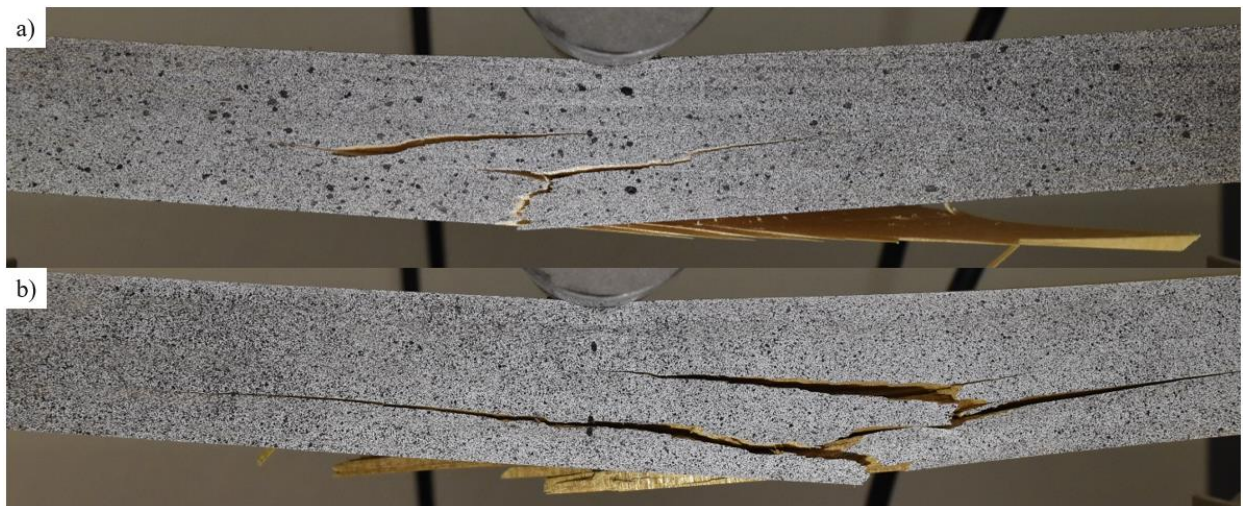

Figure 9. Typical fracture of a) UM and b) TM longitudinal flexure specimen, where the fracture path follows the grains of the specimen.

UM and TM transverse bending samples exhibited brittle, fragile failures. All samples broke suddenly, with a complete cleavage through the cross section resulting in both halves shooting off the fixture. There was no visible or audible cracking prior to ultimate failure. The fracture surface was smooth, running along the grain.

### 3.1.2 Flexural stiffness and strength

Modulus of rupture (MOR) values were calculated based on the maximum load measured in the tests. Bending strain and modulus of elasticity (MOE) were calculated based on cross head displacement of the loadframe during the elastic response of the wood prior to any crushing or other non-elastic behavior. MOE, MOR, and coefficient of variation (COV) values are presented in Table 2. As shown in Figure 10a, TM wood yielded 7% higher MOE values than UM wood for longitudinal specimens. This is consistent with the noticeably more brittle and abrupt failures exhibited by TM samples. Longitudinal MOR values were also slightly higher (+5%) in TM wood compared to UM wood (Fig. 10a). MOR and MOE values for UM and TM wood varied from those reported in the Wood Handbook (US Dept. of Agriculture 2010) for western Hemlock. Longitudinal MOE values were 24% and 32% lower for TM and UM wood respectively, but MOR values were 8% and 4% higher for TM and UM wood respectively. The variation from the properties reported in the wood handbook may be attributed to differences in grain structures, differences in density, faster tree growth, and other forest characteristics present within wood from the Makah reservation. COV values for UM and TM longitudinal bending results were consistent with a study by the FPL on mechanical properties of young growth western Hemlock collected from a similar region in Washington state (Langum, Yadama and Lowell 2009). This study reported results for flexural properties from different vertical positions within the tree and different radial sections of the tree. The trees sampled in their investigation were about 40 years younger than trees sampled for this study. They also reported lower MOR and MOE values than the wood handbook and attributed the difference to lack of maturity of the trees they used. Their reported COV values for similar sampling locations used in this study were between 9.8-15% for MOE and 10.8-12.5% for MOR. The COV values for longitudinal bending tests conducted in this study exhibited similar variability as seen in Table 2.

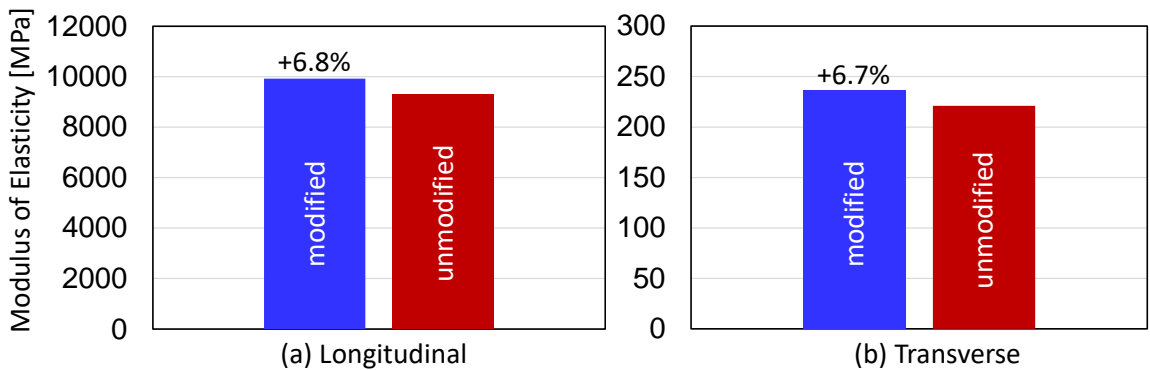

Figure 10. Comparison between the MOE of thermally modified and unmodified specimens in (a) longitudinal and (b) transverse directions.

| Orientation and sample type | Number of tests | MOR [MPa] | CoV (%) | MOE [MPa] | CoV (%) |
|---|---|---|---|---|---|
| TM Longitudinal bending | 25 | 88.4 | 11.0 | 9922 | 9.6 |
| UM Longitudinal bending | 15 | 84.0 | 12.2 | 9287 | 12.7 |
| TM Transverse bending | 13 | 3.6 | 15.5 | 236 | 14.9 |
| UM Transverse bending | 15 | 5.0 | 17.7 | 221 | 10.4 |
| Wood Handbook–West. Hem. | | 81 | | 12300 | |

Table 2: Longitudinal and transverse bending test results. Wood handbook results are taken from (US Dept. of Agriculture 2010).

For transverse bending samples there was more variation within the results. Additionally, the strength of the wood was much lower than the capacity of the load cell which adds a degree of uncertainty. As shown in Figures 10b and 11b, transverse strength of the UM wood was higher than the TM wood while transverse stiffness was slightly lower. The MOR, MOE, and COV values are presented in Table 2. Transverse bending strength is not typically studied and there are limited published results on transverse, perpendicular to grain bending strength or stiffness and no reported values in the wood handbook. The most similar mechanical property that is commonly reported is perpendicular to grain tensile strength. This is a measure of the wood's resistance to forces acting across the grain that often cause splitting (US Dept. of Agriculture 2010). Wood handbook values for perpendicular to grain tensile strength of western hemlock is 2.3 MPa. MOR values are sometimes used as conservative or low estimate of tensile strength (US Dept. of Agriculture 2010), so a higher tested MOR value than reported tensile strength could suggest a conservatively higher than average MOR in UM and TM woods tested in this study.

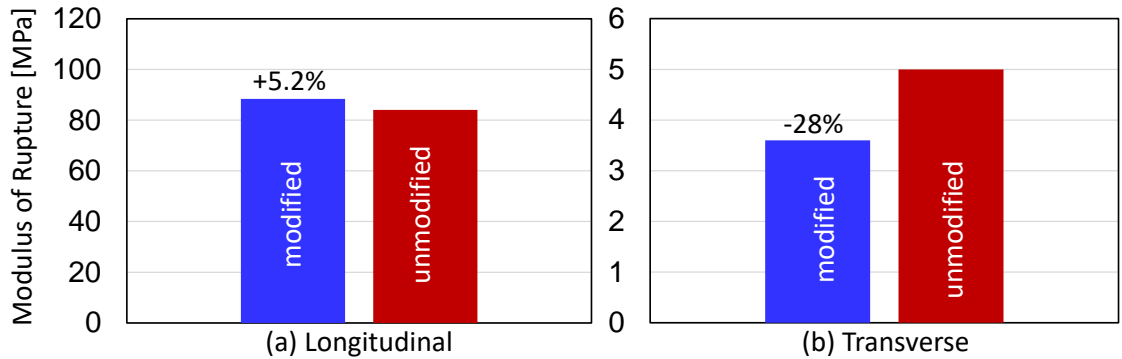

Figure 11. Comparison between the MOR of thermally modified and unmodified specimens in (a) longitudinal and (b) transverse directions.

## *3.2 Fracture tests*

Single Edge Notch Bending (SENB) specimens were tested to characterize the fracture energy of the material. The focus was on the fracturing behavior in the transverse direction which is deemed to be significantly affected by the thermal modification. As described in Figure 4, one set of specimens featured a Radial-Longitudinal (RL) configuration whereas a second set featured a Tangential-Longitudinal (TL) system. While both configurations are characterized by grains parallel to the plane of crack propagation, the different orientation of the growth ring leads to different micro/mesostructures in front of the crack tip. In turn, this leads to different damage mechanisms in the Fracture Process Zone (FPZ), leading to different energy dissipation. Since it is possible that thermal modification affects different microstructural features in different ways, it was important to characterize the fracture energy in both configurations. This also provides very useful information for the development of computational models since any anisotropic progressive damage model would require the characterization of the fracture energy in both configurations.

An interesting feature of most of the specimens tested in this work is that they exhibited stable crack propagation. This is due to the relatively high fracture energy combined with the relatively low stiffness of the samples in the transverse direction compared to the loadframe which prevented any snap-back instability (Cedolin and Z. 1991). The stable crack propagation enabled the use of the work of fracture to estimate both the initial and total fracture energy of the material (Bazant, Le and Salviato 2021). Following RILEM recommendation for mortar and concrete (RILEM 1985), the fracture energy was estimated by dividing the work of fracture $W_f$ by the ligament area. This leads to the following formula:

$$G_F = \frac{W_f}{(D-a)w} = \frac{mg\delta_0 + \int_0^{\delta_0} P d\delta}{(D-a)w} \qquad (1)$$

where, following Figure 4, $D$-$a$ is the ligament length and $w$ is the width of the specimen. Furthermore, m is the total mass of the sample, $\delta$ is the vertical displacement at the loading pin, and $g$ is the acceleration of gravity.

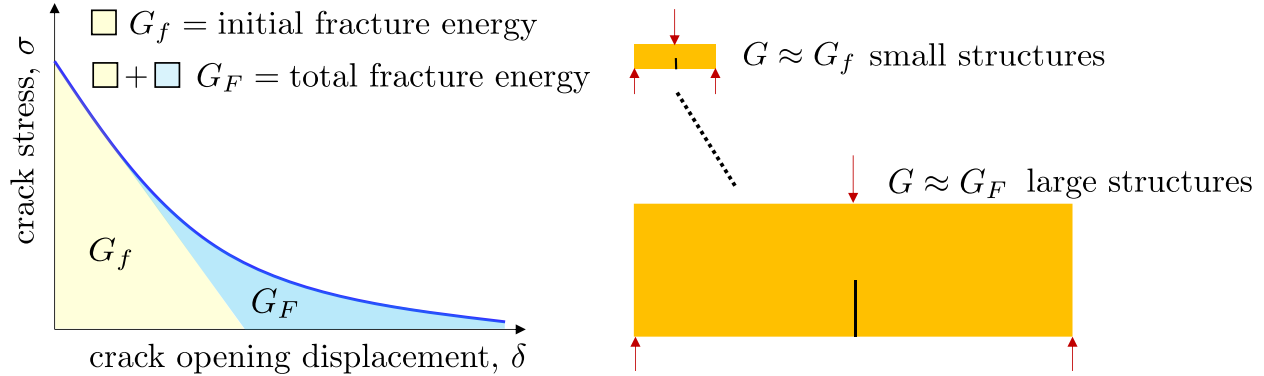

Figure 12. Typical traction-separation of a quasibrittle material (Bazant, Le and Salviato 2021) showing the initial and total fracture energies. The initial fracture energy drives the failure of relatively small structure which inhibit the full development of the Fracture Process Zone (FPZ) prior to failure. The total fracture energy drives the failure of sufficiently large structures for which the FPZ is fully developed at incipient failure.

Eq (1) is only an approximation. The main source of error is that, near the notch tip (and also near the end of the ligament), the energy to create the crack is not the same as it is in stationary propagation (stationarity is required for the fracture energy $G_F$ to be equal to the $J$-integral). However, using the work of fracture allows one to overcome a number of difficulties related to the estimation of the fracture energy from the rate of the elastic potential or using the $J$-integral (Bazant, Le and Salviato 2021). In fact, in wood samples it is very difficult to guarantee a consistent grain orientation throughout the sample, especially if the specimen is large to allow for the FPZ to develop fully. One can have a relatively good control in the region surrounding the crack but not on the whole sample. So, assuming a uniform orientation would lead to significant errors in the calculation of the energy release rate. To properly calculate the fracture energy using such approaches, a real digital twin of the sample would have to be simulated. This would make the analysis complicated and time consuming. The use of the work of fracture, on the other hand, only requires the measurement of the vertical load and vertical displacement.

Thanks to the stable crack propagation, it was possible to estimate both the initial fracture energy, $G_f$, and the total fracture energy, $G_F$ of the material. The initial fracture energy drives the failure in small samples where the FPZ cannot fully develop. It represents the area under the initial tangent of the traction-separation law of the material (Fig 11). The total fracture energy represents the total energy dissipation per crack area down to complete failure. It drives the behavior of very large notched structures for which the FPZ can develop fully prior to failure (Bazant, Le and Salviato 2021). Figure 12 shows a typical traction-separation law for quasibrittle materials exhibiting an initial and total fracture energy. $G_F$ can be accurately determined only when the load softens down to zero. This is usually difficult to achieve because either the test would have to run for a very long time or the measured reaction force would become indiscernible from the inherent noise in the testing machine. Since the measured softening curve did not extend to zero, the stress was extrapolated. The extrapolation was assumed to be an exponential decay function $Y = Ae^{-Bx}$ which is integrable up to ∞. Here $Y$= force, $x$= vertical displacement of the beam, and $A,B$= constants to be calibrated by fitting the lower portion of the softening load-displacement diagram. Figure 12 shows an example of the fitting obtained by this equation and shows the calculation of the initial and total fracture energies.

### *3.2.1 Unmodified wood*

The fracture tests of the unmodified western hemlock exhibited similar crack propagation for both RL and TL systems. In both cases, the plane of crack propagation was always orthogonal to the neutral axis of the sample, notwithstanding the difference in micro/mesotructures. In general, the fracture behavior was always relatively brittle with relatively clean fracture surfaces. Figure 13 and 14 shows the max principal strain right before crack propagation, the grain pattern at the bottom surface of the specimen, and the SEM image of the fracture surface of the RL and TL specimen respectively. The strain fields of both cases look similar, where there are strain concentrations at the crack tip and at the grains. However, the fracture surface looks different. The RL specimen exhibits damage in only one layer of the vertical tracheid cells. However, the TL specimen exhibits more damage, where about 5-10 layers of the tracheid cells fractured. Additionally, the ray cells now are in plane with the crack propagation, whereas the ray cells of the RL specimen are perpendicular. This results in delamination and fracturing of the ray cells, while the RL specimen only exhibits fracturing of the ray cells.

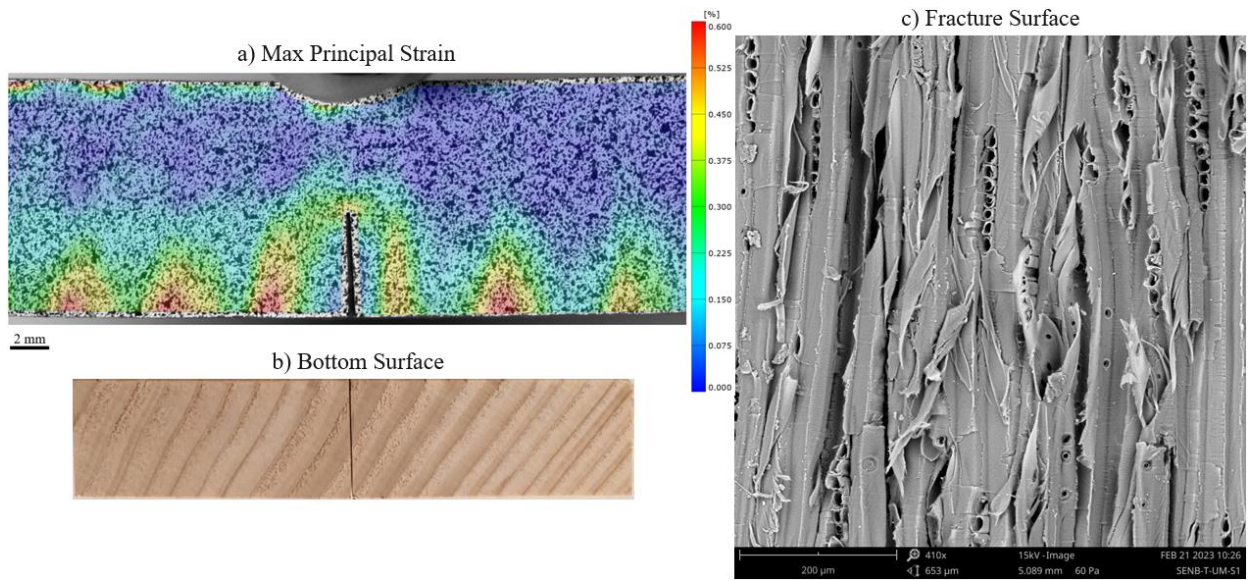

Figure 13. a) Max principal strain of the RL specimen right before crack propagation. b) Grain pattern of the specimen tested. c) SEM image of the fracture surface.

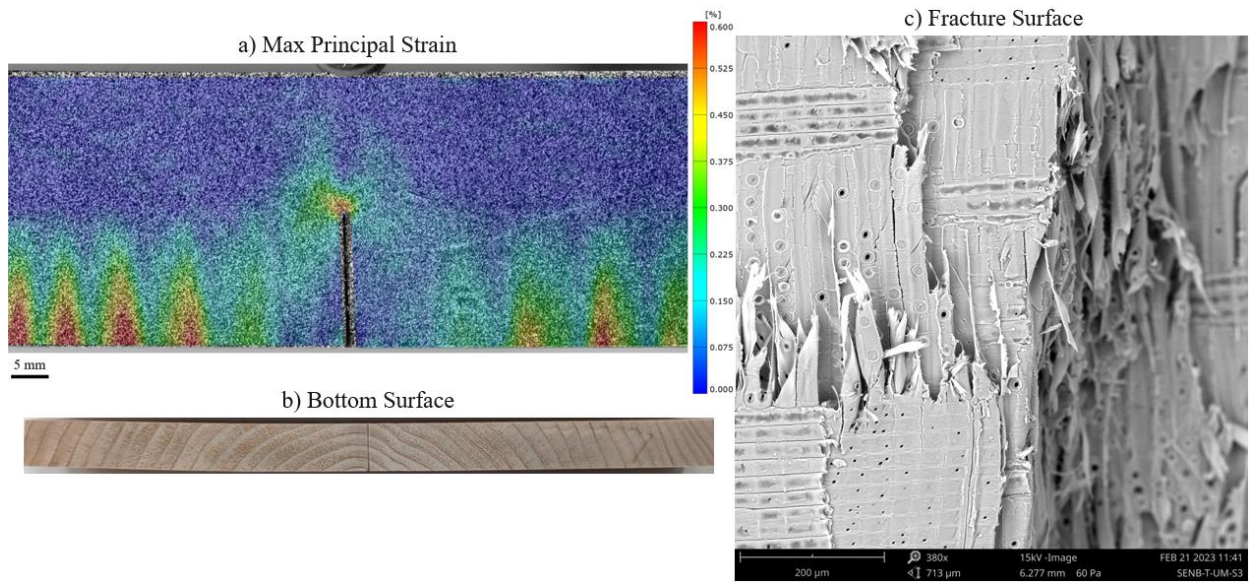

Figure 14. a) Max principal strain of the TL specimen right before crack propagation. b) Grain pattern of the specimen tested. c) SEM image of the fracture surface.

For the Radial-Longitudinal (RL) system, two specimen sizes were tested. For the small size, all the fracture tests exhibited stable crack propagation with load softening visible in the load-displacement curves. Figure 15 shows the curves measured for this configuration. The plots also show the load deflection curves after the elastic displacement is subtracted. As can be noticed, the decay function described in the foregoing section fits the shifted curves well and allows the extrapolation of the total fracture energy. For all the tests, it was possible to get a good estimation of both $G_f$ and $G_F$.

For size 2, the unmodified samples exhibited a more brittle behavior, characterized by snap-back instability for all the specimens tested (Fig. 16). In this case, the dynamic crack propagation did not allow the calculation of the fracture energy. However, this was not a problem since the results on the modified wood showed that the fracture energy calculated from Size 1 and 2 are very consistent. It is fair to assume that a similar conclusion could be made for unmodified specimens as well. This means that already for Size 1 the FPZ could develop fairly well and size effects on the fracture energy for the experiments presented in this work can be considered minimal.

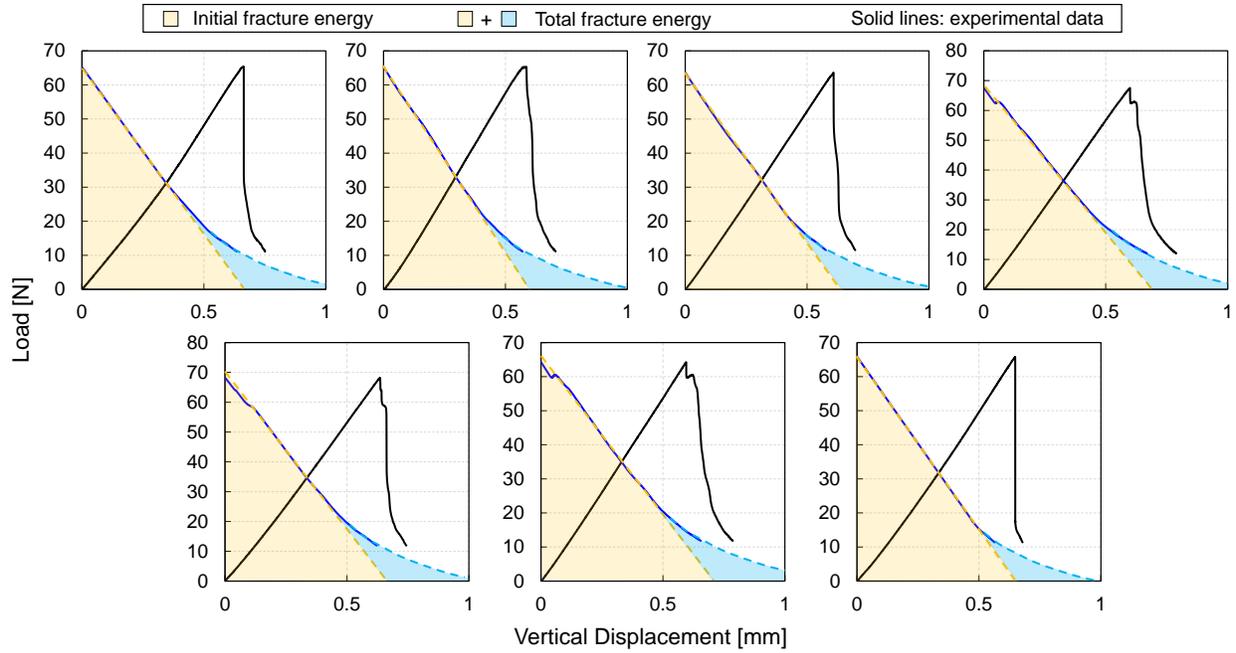

Figure 15. Load as a function of the vertical displacement for unmodified size 1 samples of the RL system. The plots also show the shifted load-displacement curves with fitted extrapolation function used to calculate the initial fracture energy and the total fracture energy.

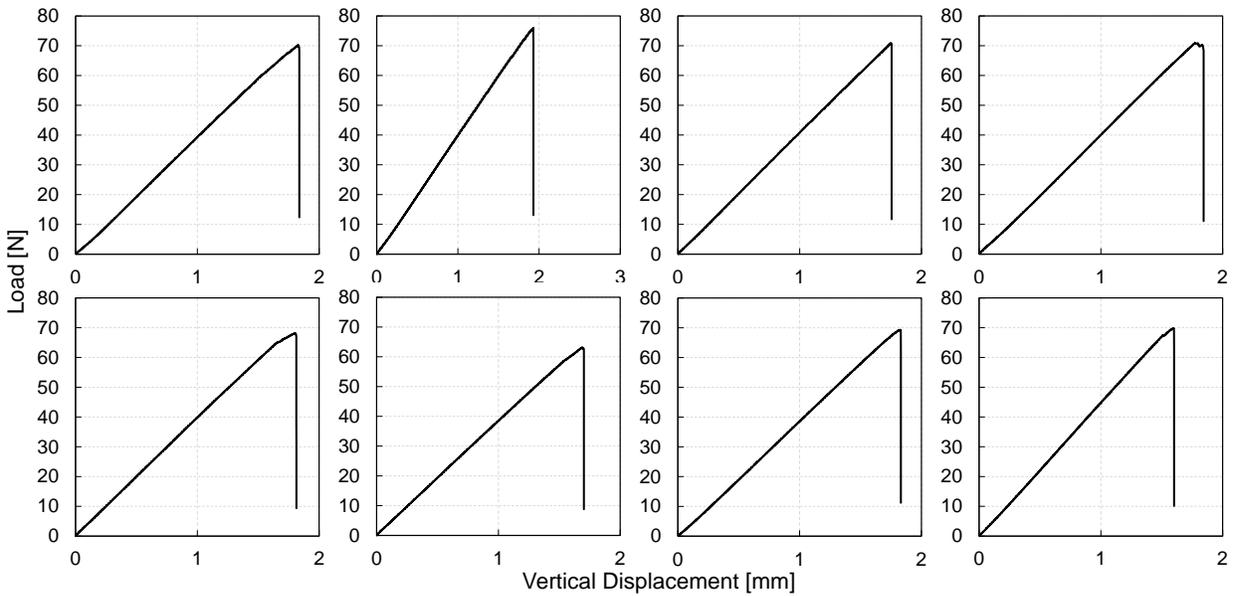

Figure 16. Load as a function of the vertical displacement for unmodified size 2 samples of the RL system. All the tests exhibited snap-back instability with dynamic crack propagation which hindered the calculation of the work of fracture.

The Tangential-Longitudinal (TL) system exhibited stable crack propagation, notwithstanding the larger size compared to both the RL systems. Fewer valid tests were available compared to the RL systems due to the presence of knots. In two samples the knots were exactly located at the crack tip effectively blunting the crack and reducing the stress intensity. Those two tests exhibited significantly larger load at failure and were discarded. A third sample featured a knot away from the crack. However, the knot was sufficiently weak that a crack initiated from it before the initial notch could start propagating. Such test was discarded as well. In any case, as Figure 17 clearly shows, the fracture tests were very consistent and allowed a full estimate of both the initial and total fracture energy for this system.

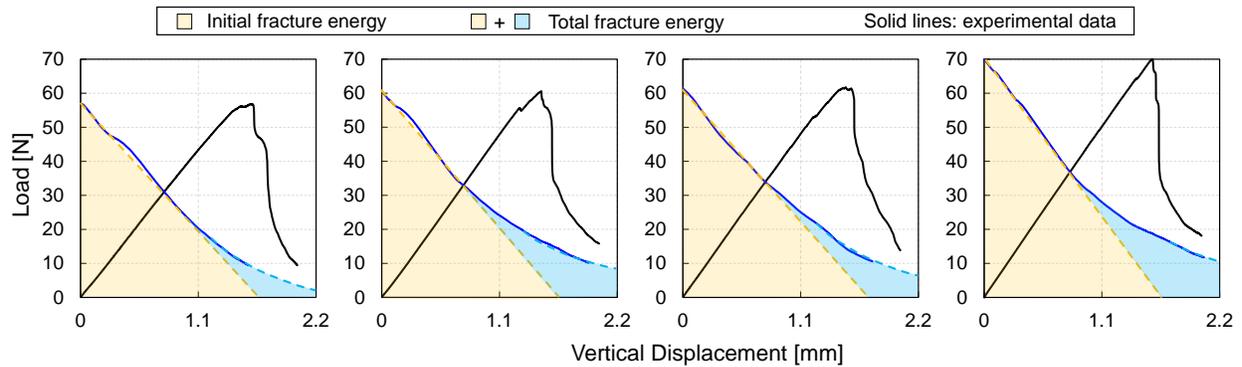

Figure 17. Load as a function of the vertical displacement for the unmodified TL system. The plots also show the shifted load-displacement curves with fitted extrapolation function used to calculate the initial fracture energy and the total fracture energy.

A summary of the fracture energies measured for the two unmodified systems is presented in Table 3. As can be noted, both the initial and total fracture energies of the RL system are significantly larger than the corresponding values for the TL system. This result is in agreement with similar tests on Spruce Wood (Murata, Watanabe and Nakano 2013). It is worth noting that the results are extremely consistent. The maximum CoV on the fracture energy is only 10.1%.

|  | RL system | | TL system | |
| --- | --- | --- | --- | --- |
|  | Mean | CoV (%) | Mean | CoV (%) |
| $G_f$ (N/mm) | 0.179 | 7.9 | 0.143 | 10.1 |
| $G_F$ (N/mm) | 0.257 | 6.4 | 0.188 | 6.4 |

Table 3. Initial and total fracture energies for both the tangential-longitudinal and radial-longitudinal configurations of unmodified wood.

### 3.2.2 Thermally modified wood

The fracture tests of the thermally modified western hemlock exhibited similar behavior to the unmodified wood. In both RL and TL systems, the plane of crack propagation was always orthogonal to the neutral axis of the sample, notwithstanding the difference in micro/mesotructures. In general, the fracture behavior was always relatively brittle with relatively clean fracture surfaces. Figure 18 and 19 shows the max principal strain right before crack propagation, the grain pattern at the bottom surface of the specimen, and the SEM image of the fracture surface of the RL and TL specimen respectively. The strain fields of both cases look similar, where there are strain concentrations at the crack tip and at the grains. However, the fracture surface looks different. The RL specimen exhibits damage in only one layer of the vertical tracheid cells. However, the TL specimen exhibits more damage, where about 2-3 layers of the tracheid cells fractured, which is much less than the UM specimen. Additionally, the ray cells now are in plane with the crack propagation, whereas the ray cells of the RL specimen are perpendicular. This results in delamination and fracturing of the ray cells, while the RL specimen only exhibits fracturing of the ray cells.

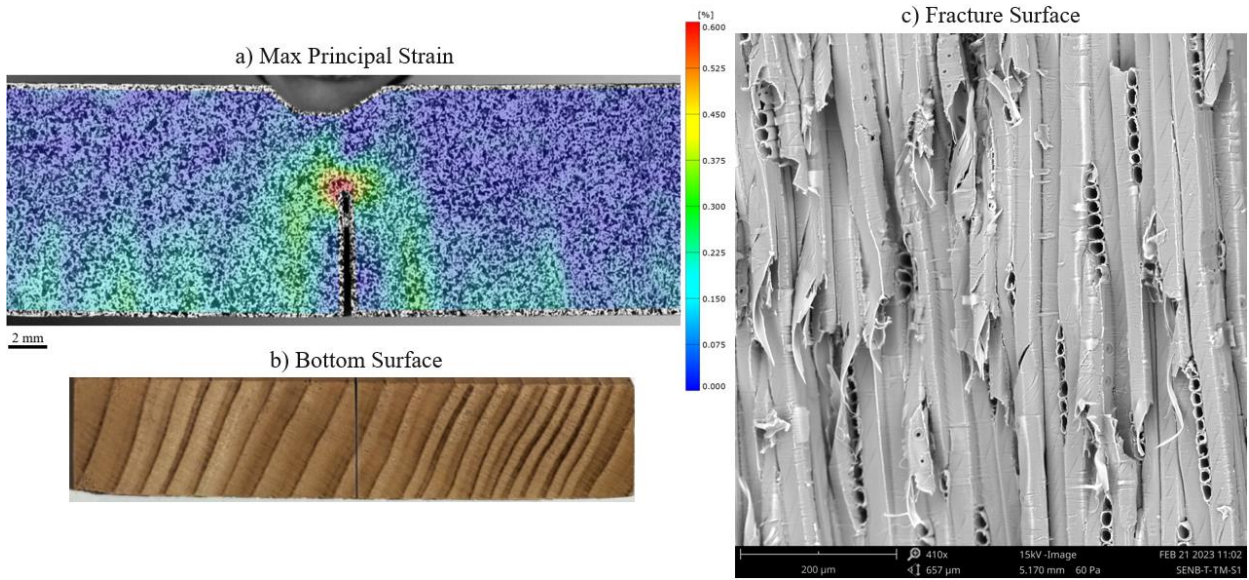

Figure 18. a) Max principal strain of the RL specimen right before crack propagation. b) Grain pattern of the specimen tested. c) SEM image of the fracture surface.

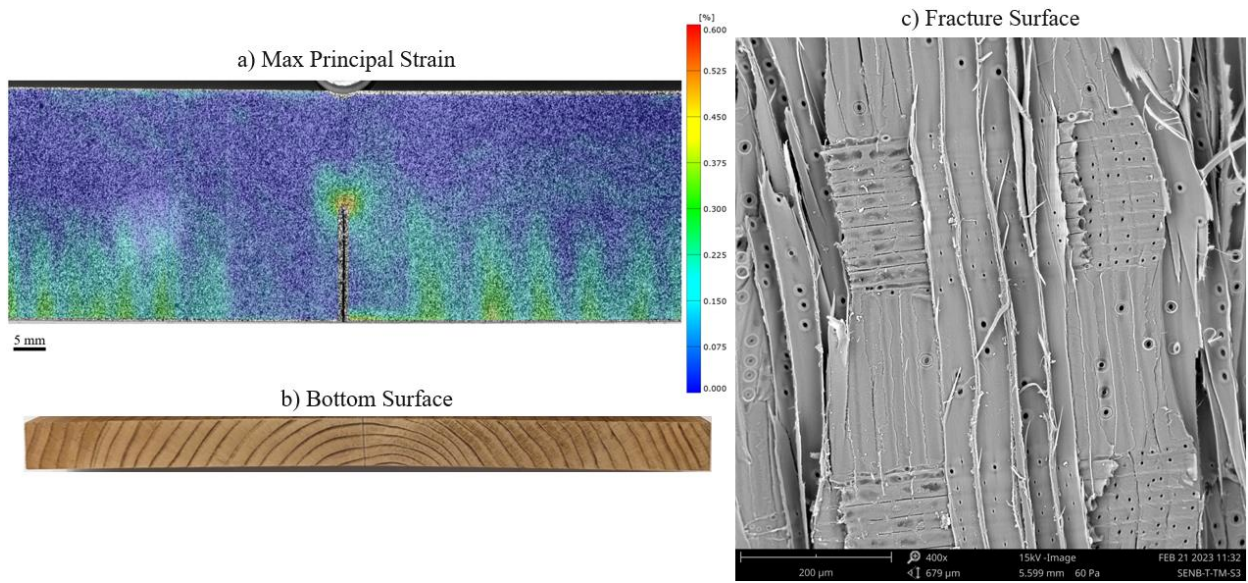

Figure 19. a) Max principal strain of the TL specimen right before crack propagation. b) Grain pattern of the specimen tested. c) SEM image of the fracture surface.

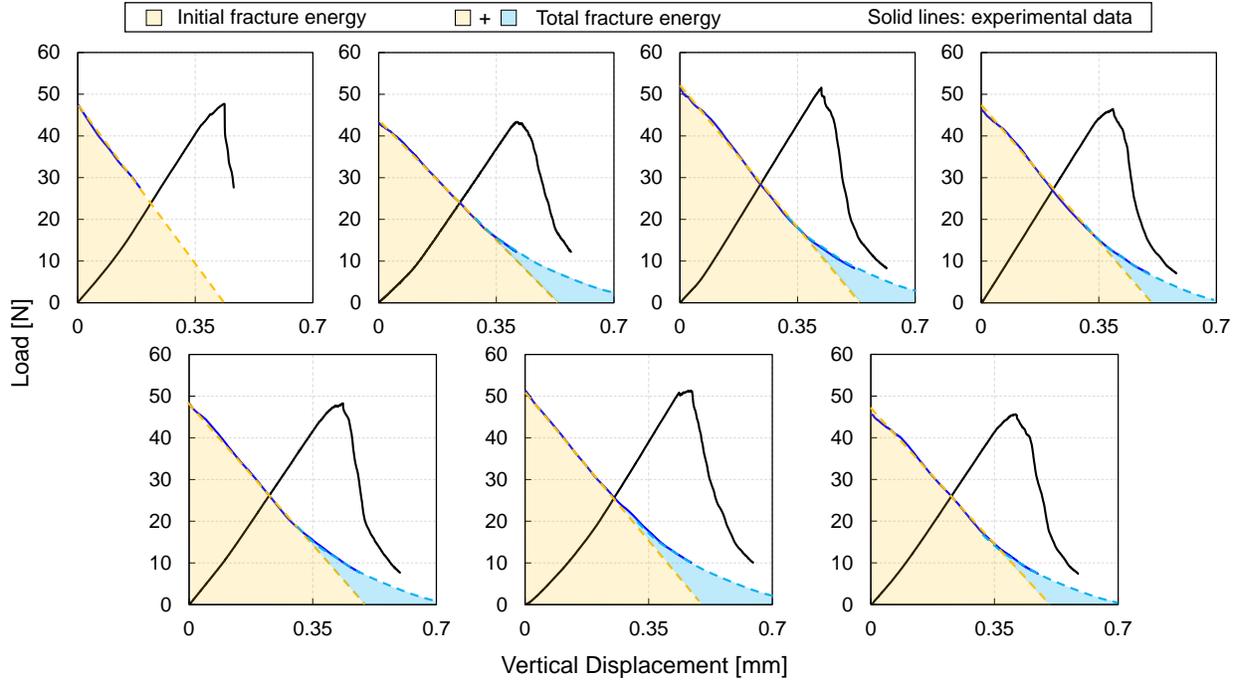

Figure 20. Load as a function of the vertical displacement for thermally modified size 1 samples of the RL system. The plots also show the shifted load-displacement curves with fitted extrapolation function used to calculate the initial fracture energy and the total fracture energy.

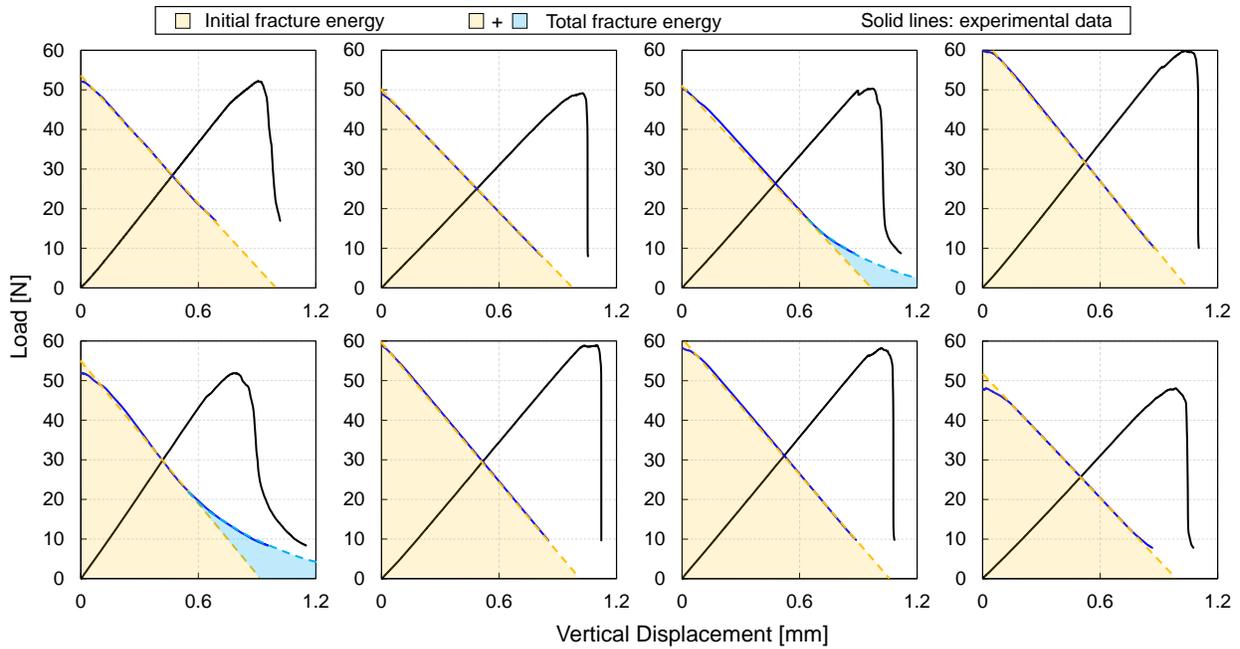

Figure 21. Load as a function of the vertical displacement for thermally modified size 2 samples of the RL system. The plots also show the shifted load-displacement curves with fitted extrapolation function used to calculate the initial fracture energy and the total fracture energy.

For the Radial-Longitudinal (RL) system, two specimen sizes were tested. For the small size, all the fracture tests exhibited stable crack propagation with load softening visible in the load-displacement curves. All the tests but one allowed the characterization of both the total and initial fracture energies (Figure 20). For the larger size, crack propagation was rather stable although only two tests allowed the full characterization of the total fracture energy (Fig. 21).

The Tangential-Longitudinal (TL) system displayed a similar behavior to the unmodified wood. For all the tests, crack propagation was stable. As can be noted from Fig. 22 the load-displacement curves were very consistent and allowed a fairly accurate calculation of the fracture energies.

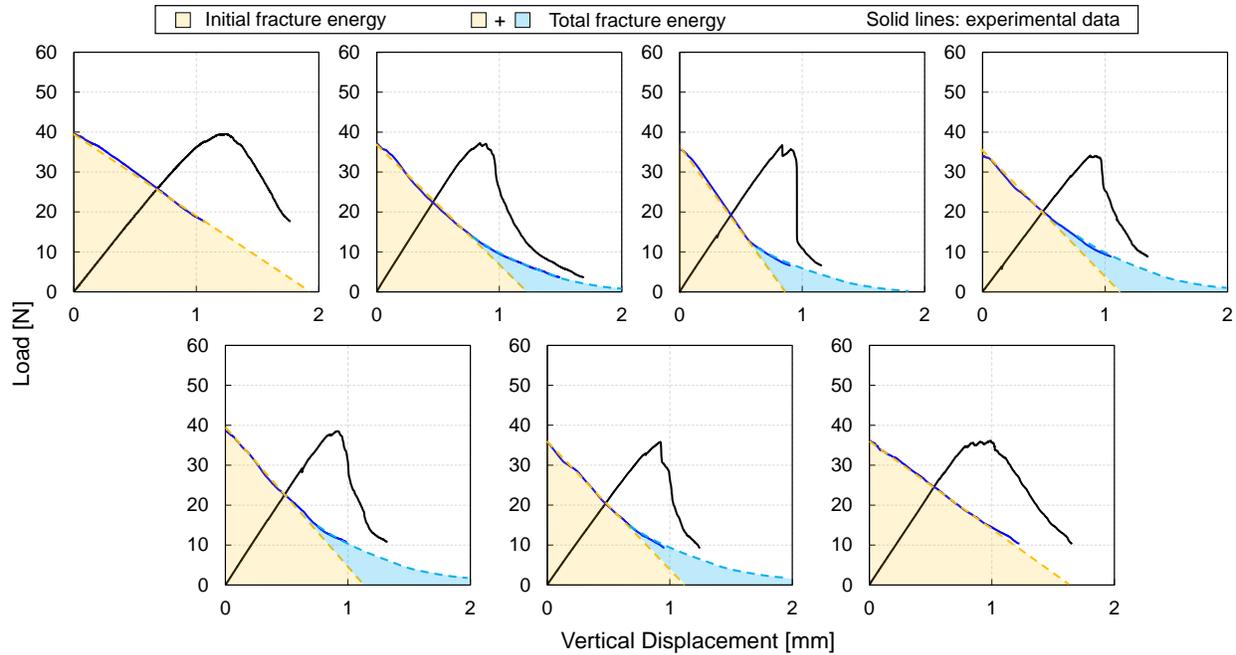

Figure 22. Load as a function of the vertical displacement for the thermally modified TL system. The plots also show the shifted load-displacement curves with fitted extrapolation function used to calculate the initial fracture energy and the total fracture energy.

Table 4 presents a summary of the fracture energies of the thermally modified wood. As can be noted, even in this case the RL system features a larger fracture energy compared to the TL system, consistent with the unmodified wood. However, the values of the fracture energies are significantly lower compared as it will discussed next.

|  | RL system | | TL system | |
| --- | --- | --- | --- | --- |
|  | Mean | CoV (%) | Mean | CoV (%) |
| $G_f$ (N/mm) | 0.106 | 11.8 | 0.069 | 27.3 |
| $G_F$ (N/mm) | 0.136 | 9.1 | 0.075 | 18.4 |

Table 4. Initial and total fracture energies for both the tangential-longitudinal and radial-longitudinal configurations of thermally modified wood.

*3.2.3 Comparison between thermally modified and unmodified wood*

The accurate calculations of the initial and total fracture energies presented in the foregoing sections give an opportunity to evaluate the effect of thermal modification on these important properties. As shown by Figure 23, thermal modification leads to quite a significant embrittlement in the RL system. The initial fracture energy is reduced by 41% while the total fracture energy is reduced by 47%.

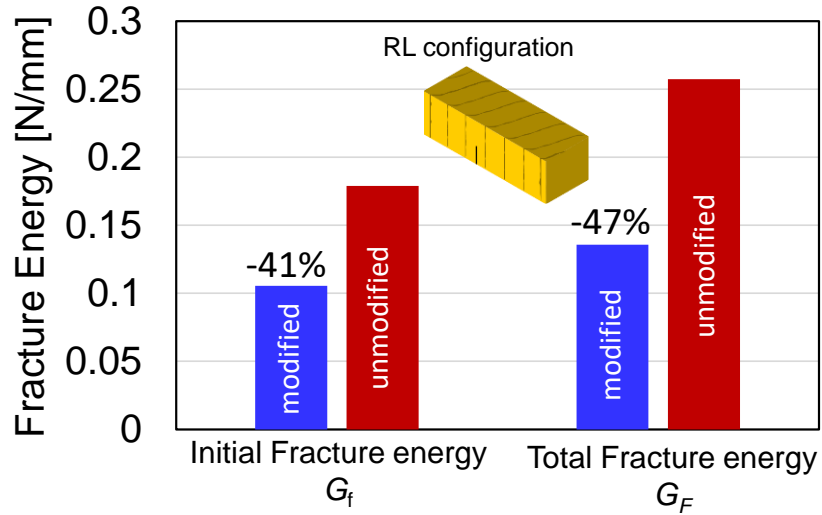

Figure 23. Comparison between the initial and total fracture energy of modified and unmodified wood for the Radial-Longitudinal (RL) configuration

Similar conclusions can be drawn for the TL system. As Figure 24 shows, in this case the reduction is even more dramatic: 52% for the initial fracture energy and 60% for the total fracture energy.

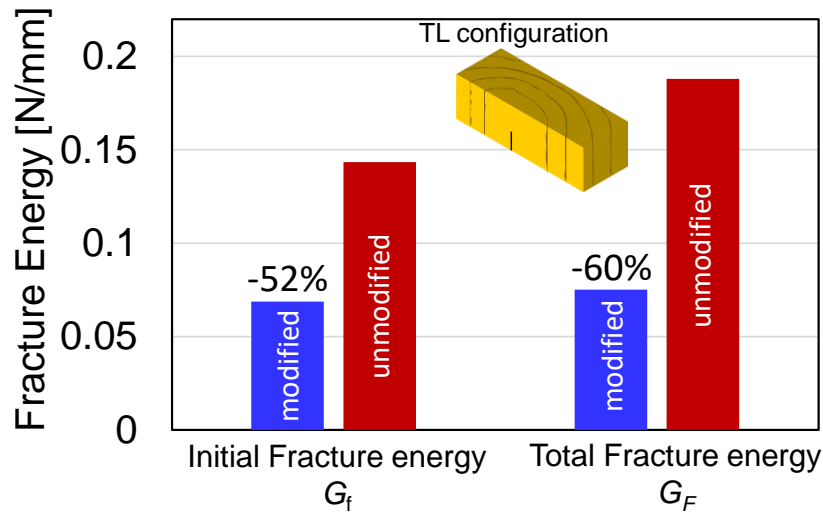

Figure 24. Comparison between the initial and total fracture energy of modified and unmodified wood for the Tangential-Longitudinal (TL) configuration

There are a few factors that could be causing the change in fracture energy, 1) change in moisture content, 2) change in degree of polymerization, and 3) degradation of the hemicellulose in the wood. Previous work on other wood species, on cellulose nanocrystal films, and other polymer composites have shown that the moisture content has a significant effect on the mechanical properties of the material (Hou, et al. 2020, Arnold 2010, Wang and Wang 1999, Nakagawa, et al. 2022). These works showed that the water molecules can act as a plasticizer, allowing for the polymer chains to slide more easily and increasing energy absorption. It has been seen in nanocellulose films, that the degree of polymerization of the cellulose has an impact on the strength of the material (Fang, et al. 2020). However, with wood, the change in the degree of polymerization of the cellulose is species dependent and may have a limited change at the low temperature thermal modification done in this work (Kubovský, Kačíková and Kačík 2020). Finally, the degradation of wood due to thermal and chemical modification has been studied previously and has shown that the degree of degradation is species and process dependent (Sikora, et al. 2018, LeVan, Ross and Winandy 1990). However, when degradation is significant, it is typically seen in the hemicellulose. This degradation would cause a poor transfer of stress between the cells and cause a decrease in fracture energy.

### *3.3 Janka hardness tests*

Figure 25 and Table 5 contains the results of the Janka hardness tests for each configuration and wood face. The average hardness values are higher in the UM specimen for all wood faces and thermal modification ultimately caused a statistically significant decrease in surface hardness of the tangential and transverse wood planes. These findings are consistent with findings from previous work, that surface hardness generally decreased with thermal modification, especially at increased temperatures (Nourian and Avramidis 2021, S. Nourian 2018). Although the TM specimens in this study underwent minimum TM temperatures, the difference is still apparent. Clearly, the effect and degree of thermal modification on Western hemlock hardness must be considered when designing with this wood species.

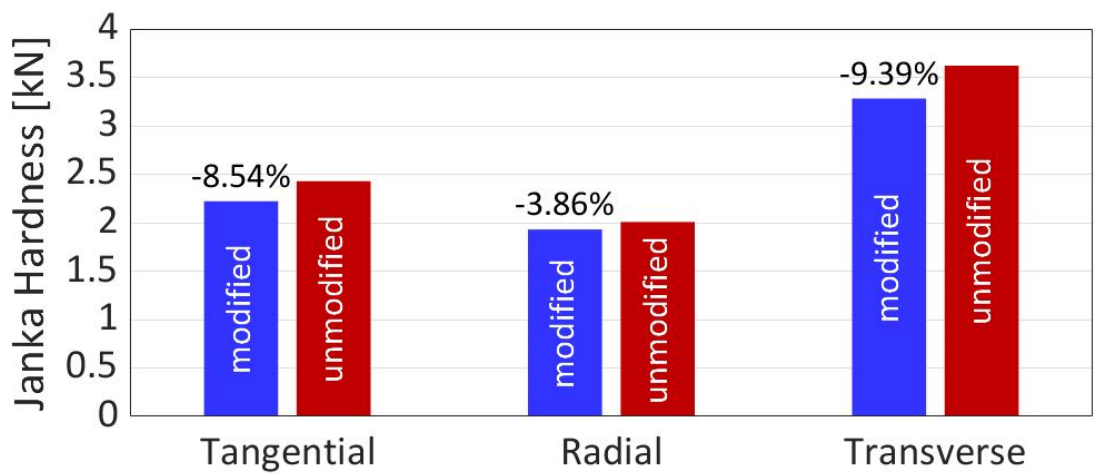

Figure 25: Janka hardness values for Western hemlock.

|  | UM system | | TM system | |
| --- | --- | --- | --- | --- |
|  | Mean [kN] | CoV (%) | Mean [kN] | CoV (%) |
| Tangential | 2.43 | 14.95 | 2.22 | 13.10 |
| Radial | 2.01 | 17.55 | 1.93 | 14.50 |
| Transverse | 3.62 | 9.79 | 3.28 | 13.44 |

Table 5. Janka hardness of UM and TM Western Hemlock on the tangential, radial, and transverse faces.

## 4 Conclusions

This article investigated the effect of thermal modification on the longitudinal and transverse flexure properties, fracture energy, and hardness of Western Hemlock. Based on the results of this work, the following conclusions can be elaborated:

1. Thermal modification has significant effects on the mechanical properties of Western Hemlock, especially in the direction transverse to the grains;
2. In the longitudinal direction, thermal modification led to slight increases in Modulus of Elasticity (MOE) and Modulus of Rupture (MOR). In fact, the MOE increased by 6.8% while the MOR increased by 5.2%;
3. In the transverse direction, thermal modification led to slight increases in Modulus of Elasticity (MOE), which increased by 6.7%. However, the transverse MOR saw a significant drop by 28%;
4. The fracture energy of both Radial-Longitudinal (RL) and Tangential-Longitudinal (TL) systems showed a dramatic reduction due to thermal modification. In the RL system, the initial and total fracture energies decreased by 41% and 47% respectively. For the TL system, the initial and total fracture energies decreased by 52% and 60%;
5. The Janka hardness values were reduced on average due to the thermal modification treatment and this reduction was statistically significant for the tangential and transverse wood planes.

The foregoing results are very important for the design of western hemlock structures. The reduction of the fracture energy and Janka hardness induced by thermal modification must be taken into serious consideration since they correlate strongly to the capacity of the material to be damage tolerant and to dissipate energy upon crushing. Ballistic performance is known to be strongly correlated to these properties as well.

## 5 Acknowledgments

This material is based upon work supported by the US Army Engineer Research and Development Center (ERDC) under contract W9132T22C0008. The tests described and the resulting data presented herein, unless otherwise noted, are supported under PE 0603119A, Project BO3 'Military Engineering Technology Demonstration (CA)', Task 'Program Increase - Cross-Laminated Timber and Recycled Carbon Fiber Materials'.